# A Secure Database System using Homomorphic Encryption Schemes

[1]Youssef Gahi, [2]Mouhcine Guennoun, [2]Khalil El-Khatib

[1]Ecole Mohammadia d'Ingénieurs B.P 765 Avenue Ibn Sina, Agdal,
Rabat, Morocco
[2]University of Ontario Institute of Technology, 2000 Simcoe Street North,
Oshawa, Ontario, Canada. L1H 7K4
youssef.gahi@gmail.com, mouhcine.guennoun@uoit.ca, khalil.el-khatib@uoit.ca

*Abstract*—Cloud computing emerges as an attractive solution that can be delegated to store and process confidential data. However, several security risks are encountered with such a system as the securely encrypted data should be decrypted before processing them. Therefore, the decrypted data is susceptible to reading and alterations. As a result, processing *encrypted* data has been a research subject since the publication of the RSA encryption scheme in 1978. In this paper we present a relational database system based on homomorphic encryption schemes to preserve the integrity and confidentiality of the data. Our system executes SQL queries over encrypted data. We tested our system with a recently developed homomorphic scheme that enables the execution of arithmetic operations on ciphertexts. We show that the proposed system performs accurate SQL operations, yet its performance discourages a practical implementation of this system.

## I. INTRODUCTION

Cloud computing is an attractive solution that can provide low cost storage and processing capabilities for government agencies, hospitals, and small and medium enterprises. It has the advantage of reducing the IT costs and providing more services for the requesting parties through making specialized software and computing resources available. However, there are major concerns that should be considered by any organization migrating to cloud computing. The confidentiality of information as well as the liability for incidents affecting the infrastructure arise as two important examples in this context. Indeed, cloud computing poses several data protection risks for the cloud's clients and providers. For example, the cloud's client may not be aware of the practices according to which the cloud's provider processes the stored data. Therefore, the cloud's client cannot guarantee that the data are processed (for example, altered or deleted) in a legal and accepted manner.

All of the above mentioned issues can be resolved if the data in the cloud are stored and processed in encrypted form. The latter is possible if the encryption scheme can support *addition* and *multiplication* of the encrypted data. Many encryption schemes support one of these operations, like the encryption schemes in [1-4]. A cryptosystem which supports both addition and multiplication (referred to as the homomorphic encryption scheme) can be effective data protection, and enables the construction of programs that receive encrypted input and produce encrypted output. Since such programs do not decrypt the input, they can be run by an un-trusted party without revealing their data and internal states. Such programs will have great practical implications in the outsourcing of private computations, especially in the context of cloud computing.

Homomorphic cryptosystems have received valuable attention in the literature, see [5][6][7]. In theory, the data can be encrypted by the client, and then sent to the cloud's provider for storage or processing. Only the client holds the decryption keys necessary to read the data. Despite the fact that this type of processing may increase the amount of computing time, the benefits associated with it are worth the processing overhead. Indeed, this model of computing can preserve the confidentiality and integrity of the data while delegating the storage and processing to an un-trusted third party.

In this paper, we present a novel technique to execute SQL statements over encrypted data. We develop a secure database system that processes these queries. The parameters of SQL queries are encrypted by the client and sent to the server for processing. The latter performs the requested operation over an encrypted database and returns an encrypted result to the client. The advantage of this system is that the database server knows neither the content nor the position of the records affected by the query.

The remainder of this paper is organized as follows. In Section II, we review the literature for the work related to private information retrieval (PIR) approaches. Section III provides a formal description of our secured SQL statements approach. Section IV presents a homomorphic cryptosystem that we use to build a prototype system. Section V presents an implementation of a secure relational database system. In Section VI we provide performance analysis of the proposed secure database system. Finally, Section VII concludes our work and provides future research directions.

## II. PRIVATE INFORMATION RETRIEVAL

*Chow et al.* [8] discussed the importance of cloud computing, and how this technology can be enticing due to its flexibility and cost-efficiency. The authors pointed out that the adoption of such technology is still below ambition. Some users are still concerned about the security of these clouds. Even those who started using the technology, they only utilize it with their less sensitive data. The limited usage of cloud computing is mainly due to the lack of control over the communicated data. The authors highlight that people require explicit guarantees that their data will be protected under well-defined policies and mechanisms. However, no technical security solutions were proposed to back-up their *information centric* model where data can self defend itself in a hostile or an un-trusted environment.

The private information retrieval (PIR) approach, introduced by *Chor et al.* in [9], achieves the retrieval of an $i^{th}$ bit in a block without revealing information about the bit retrieved or about the request for the bit itself. This approach has been widely used as a basis for several tools, and has supported various distributed applications. However, the approach requires more improvements and the work with it is still in progress, both at the security of the communication channel level and the hidden client identity level.

*Raykova et al.* [10] extended the PIR approach by proposing a secure anonymous search system. The system employs keyword search such that only authorized clients have access to their blocks. This system is capable of mapping the database content to the appropriate client, thus guaranteeing the privacy of the data and the query. The ultimate target of *Raykova*'s system is to ignore the identity of the client while protecting the database from malicious queriers.

*Shang et al.* [11] tackled the problem of protecting the database itself. The problem is studied through monitoring the amount of data disclosed by a PIR protocol during a single run. The information attained from the monitoring process is used to understand how a malicious querier can conduct attacks to retrieve excessive amount of data from the server.

PIR has also been used to develop authentication systems. *Nakamura et al.* [12] constructed a system with three components, a querier that initiates requests, an authentication-server that processes these requests, and a database that returns the appropriate data in response to the request. This system ensures the security of data and the anonymous communication between the querier and the database. *Yinan and Cao* [13] used the PIR approach to propose a system that controls the access to the database. According to this system, the privacy of data is enforced by enabling each authorizer to give or deny access to his/her own data with a hierarchical authorization access right scheme.

Among the most important criteria in PIR protocol are the communication cost and the amount of data sent back to the querier. The trivial solution of the PIR protocol is to send back the entire database to the client. However, this solution is expensive, even for a simple request that results in retrieving two matching records. Other approaches proposed to retrieve only the requested data, by using replicated databases that are stored at multiple servers. In this case, the request is forwarded to all servers. With this approach, although we deal with multiple replicated databases, the privacy is better protected. However, this approach is still complicated and may result in extended processing and communication times. *Gentry et al.* [14] proposed a scheme to retrieve a bit or a block from a database with a constant communication rate. *Melchor et al.* [15] proposed a scheme that reaches the available data with a reasonable communication cost while achieving lower computational cost compared to other PIR protocols.

## III. SECURED SQL OPERATIONS

In this section, we develop a secure database system that processes SQL queries over encrypted data. As shown in Figure 1, parameters of the queries are encrypted by the client and sent to the server for processing. The latter performs the requested operation and returns encrypted results to the client.

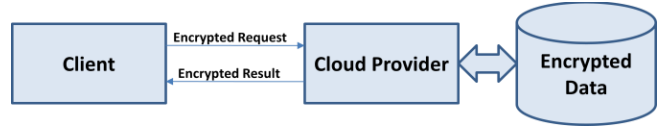

Figure 1. Secure Data Retrieval

We describe below the circuit of a simple SQL SELECT query:

```
SELECT * from T where c=v
```

where the value *v* is in encrypted form. The trivial solution to securely perform this statement is to send back to the client the entire database, but this solution suffers from complexity and scalability issues. Instead, we propose a methodology to implement the SELECT circuit at the server side, while preserving the confidentiality and the privacy of the request.

The processing of the SELECT query is divided into three sub-circuits. Firstly, we calculate the following index for each record *R* in the table *T*:

$$\forall R \in T \; I_R = \prod_{i=0}^{size-1} (1 \oplus c_i \oplus v_i)$$

where *size* is the number of bits in column *c*; $c_i$ and $v_i$ are the $i^{th}$ bits of column *c* and search criteria *v*, respectively. $I_R$ is a one bit value that is equal to 1 if *v* matches the value of column *c*, 0 otherwise.

Next, we identify the $n^{th}$ record that matches the selection criteria. For that purpose, we consider $\eta = \varepsilon_{pk}(n)$ to be the encryption of *n* under public key *pk*.

For each record *R* we calculate the following sum:

$$\forall R \in T: S_R = \sum_{i \leq R} I_R$$

We calculate a second index $I'_R$:

$$\forall R \in T \; I'_R = I_R \times \prod_{i=0}^{size-1} (1 \oplus \eta_i \oplus S_{R,i})$$

$I'_R$ is equal to 1 if the record *R* is the $n^{th}$ record that matches the selection criteria, 0 otherwise.

Then, we multiply every bit of each record *R* in table T by the corresponding value $I'_R$.

$$\forall R \in T: R' = I'_R \times R$$

This latter operation forms a table $T'$ that is related to the original table $T$ as follows:

$$\begin{cases} R' = R \text{ if } R \text{ is the } n^{th} \text{ record matching the criteria} \\ R' = 0 \text{ otherewise} \end{cases}$$

Finally, by adding all records of table $T'$, we retrieve the $n^{th}$ record $R_s$ that matches the selection criteria:

$$R_s = \sum_{R' \in T'} R'$$

If no record matches the selection criteria, a record containing zeros will be returned to the requester.

It is worth noting that all calculations are performed over encrypted data. The server does a blind processing to retrieve the $n^{th}$ record that matches the selection criteria. It neither has access to the content of the retrieved record nor to its position within the table.

With slight modifications to the select circuit, most of SQL operations can be supported by our proposed secure database system. For example, to implement the UPDATE operation, one can simply implement the following circuit:

$$\forall R \in T: R' = \overline{I_R} * R + I_R * U$$

where the record $U$ is the new value to update the record $R$ matching the criteria of the query.

Similarly, to delete a record from table $T$, one can replace its content by zero. The DELETE operation can be implemented by the following circuit:

$$\forall R \in T: R' = \overline{I_R} * R$$

## IV. HOMOMORPHIC ENCRYPTION SCHEME

In [6], Gentry proposed a fully homomorphic encryption scheme that enables to perform an arbitrary number of arithmetic operations (i.e. addition and multiplication) on encrypted data. The components of the encryption scheme are described below.

Security Parameters: $N = \lambda$, $P = \lambda^2$, and $Q = \lambda^5$.

### A. Key Generation

The private key $s_k$ is a random P-bit odd number. The public key consists of a list of integers that are the "encryptions of zero" using the encryption scheme with the secret key $s_k$ as a public key.

Generate a set $\vec{y} = \{y_1, ..., y_\beta\}$ of rational numbers in [0,2[ such that there is a sparse subset $S \in \{1, ..., \beta\}$ of size $\propto$ with $\sum_{i \in S} y_i \approx 1/p \mod 2$.

Set $sk^*$ to be the sparse subset S, encoded as a vector $s \in \{0,1\}^\beta$ with hamming weight $\alpha$.

Set $pk^* \leftarrow (pk, \vec{y})$ to be the public key.

### B. Encryption (pk*,m)

Set $m'$ to be a random N-bit number such that $m$ and $m'$ have the same parity:

$$m' = m\%2$$

Then compute $c$ as:

$$c \leftarrow m' + pq$$

where $q$ is a random Q-bit number. Then the ciphertext $c$ is post-processed to produce a vector $\vec{z} = \{z_1, ..., z_\beta\}$, defined by:

$$z_i \leftarrow c.y_i \mod 2$$

The output ciphertext $c^*$ consists of $c$ and $\vec{z} = \{z_1, ..., z_\beta\}$.

### C. Decryption (sk*, c*)

$$m \leftarrow LSB(c) \oplus LSB(\sum_i s_i.z_i)$$

### D. Arithmetic Operations

Addition and multiplication can be performed on clear text by simply adding and multiplying the ciphertexts, respectively.

$$\varepsilon_{pk}(m_1 * m_2) = \varepsilon_{pk}(m_1) * \varepsilon_{pk}(m_2)$$

$$\varepsilon_{pk}(m_1 + m_2) = \varepsilon_{pk}(m_1) + \varepsilon_{pk}(m_2)$$

The output ciphertext $c^*$ consists of $c$ together with the result of post-processing the resulting ciphertext with $\vec{y}$.

### E. Bootstrapping the Encryption Scheme

The scheme described above is referred to as a *somewhat* homomorphic scheme because it works only if the value $c\%p$ (noise of the encryption) is smaller than $p/2$. After a finite number of arithmetic operations, the noise exceeds the $p/2$ threshold and the decryption scheme does not work anymore.

Gentry developed a novel method to remove the noise in the ciphertext [7]. He proposed to recrypt the ciphertext $c$ to remove the noise. Since the scheme is homomorphic, one can encrypt the ciphertext $c$ into a new ciphertext $\bar{c}$ (the plaintext is encrypted twice), and by using the homomorphic properties of the scheme, one can decrypt the inner layer of encryption to obtain a ciphertext $c_2$ with a lower value of noise.

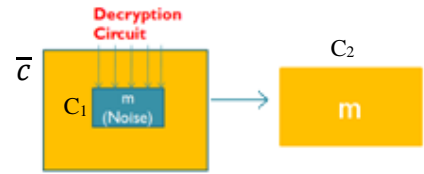

Figure 2. Removing noise from original ciphertext (bootstrapping)

As illustrated in Figure 2, a bit $m$ is encrypted with public key $pk$ to produce the ciphertext $c_1$. After a finite number of arithmetic operations, the noise associated with the ciphertext $c_1$ reaches a level that does not permit any additional arithmetic operation. To remove the noise, the bootstrapping technique consists of recrypting the bit $m$. Every bit of ciphertext $c_1$ is encrypted with the public key $pk$. The output is ciphertext $\bar{c}$ that doubly encrypts bit $m$. The decryption circuit is applied to remove the inner layer of encryption. This latter operation requires the knowledge of the private $sk$. Therefore, the private key is encrypted with public key $pk$; and then shared with the server. Since the encryption scheme is homomorphic, the decryption can be performed on the doubly encrypted ciphertext to remove the inner layer. The recryption produces a new ciphertext $c_2$ with a value of noise that has an upper bound according to the proof in [6].

By employing the bootstrapping technique, performing an arbitrary number of arithmetic operations on ciphertexts becomes possible.

## V. IMPLEMENTATION

In our implementation, we aim at proving that it is possible to perform SQL queries over an encrypted database. For example, the user can specify a search criterion through a database. Then, the client software encrypts the parameters of the query, corresponding to the search criterion, and sends it to the appropriate server. The server retrieves the requested record (blind processing) from the database and returns it to the client. The client software decrypts the record and displays it to the user.

We built a simple medical application containing 10 patients' records. In Figure 3, we show the result of the SELECT query. This is how the result appears in a screenshot of the client side of our built application.

The application supports the following SQL operations:
- SELECT with wildcard characters (*, ?) and relational operators (< >).
- UPDATE with wildcard characters (*, ?) and relational operators (< >).
- DELETE with wildcard characters (*, ?) and relational operators (< >).
- Statistical operations like COUNT and AVG.

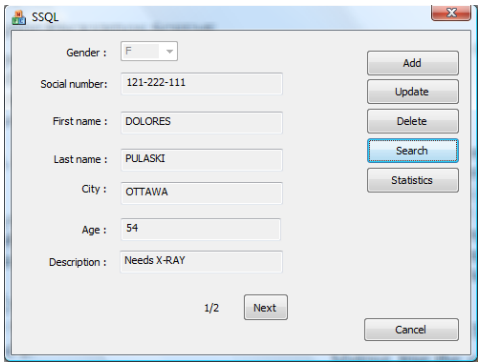

Figure 3. Client side of the application

It is worth mentioning that the implementation of the medical application was built using a simplified and non-secure version of the somewhat homomorphic scheme. This is due to performance issues as it is impractical to perform our tests using the fully homomorphic cryptosystem. We chose the security parameters in such a way to support all the SQL operations with no need to employ the bootstrapping technique. We discuss the performance of our system in the next section.

## VI. PERFORMANCE ANALYSIS

Table 1 lists the number of arithmetic operations required to execute some basic SQL statements over an encrypted database of 10 records. From this table we can see that processing data in encrypted form creates a substantial computation overhead.

TABLE I. NUMBER OF ARITHMETIC OPERATIONS

|  | Add. & Mult. | Add. | Mult. |
| --- | --- | --- | --- |
| SELECT | 619839 | 309892 | 309947 |
| UPDATE | 67595 | 25355 | 42240 |
| DELETE | 28171 | 5643 | 22528 |

To understand the processing time required to process a SQL statement, we measured the time required to perform the product of two n-bits numbers in encrypted form using the fully homomorphic cryptosystem presented in [6]. Towards that end, we used a computer machine with 1.7 GHz processor and 3GB of RAM memory. Figure 3 shows the amount of time, in seconds, required to compute the multiplication circuit.

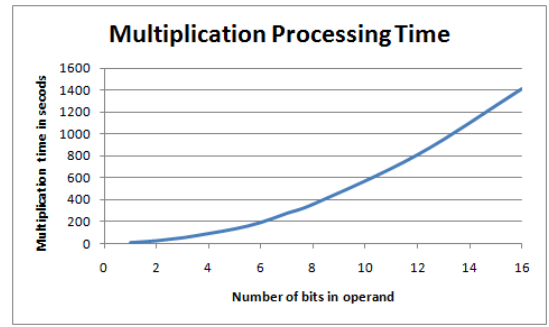

Figure 4. Processing time required to perform the product of two n-bits integers

As we can see in Figure 4, it takes 23 minutes to compute the product of two 16-bit integers. This latency is due to the bootstrapping technique or more precisely to the recrypt function. Indeed, according to our measurements, it takes 1 second to recrypt a ciphertext. Therefore, there is a need of at least 7 days (i.e, 619839 * 1 second) to retrieve a row from a 10-record database.

The implementation of the system proves that the execution of SQL statements over encrypted data is feasible.

However, the time required to execute these statements is very high and therefore is not suitable for real-time transactions that involve a large database (i.e. several terabytes database). This drawback is mainly due to the homomorphic encryption scheme. In fact, there might be more efficient techniques to optimize the implementation, that is, one could perform recryption only when it is necessary, since the noise value can be bounded; however, we do believe that a more practical homomorphic cryptosystem is yet to be developed.

## VII. Conclusion & Future Directions

The concept of processing encrypted data is promising to revolutionize traditional computing. Indeed, this concept has many direct applications in cloud computing environments, banking, electronic voting and many other applications.

In this paper we developed the first secure database system based on a fully homomorphic encryption scheme. We presented the circuits to implement SQL statements over encrypted data. We built a prototype of a database system where data is stored and processed in encrypted form. The database server can execute most of the SQL statements in a blind fashion, that is, it returns the results without any knowledge of the content or the position of the records extracted/affected. We conducted performance analysis to measure the time needed to execute a simple query on the database. We found that the current technology is not sufficiently mature yet as it is time-consuming. Indeed, the encryption schemes proposed by *Gentry et al.* in [5][6][7] are very impractical. According to our measurements, the time needed to perform simple calculations is substantial. We believe that there still is a great opportunity for researchers to develop more efficient homomorphic encryption schemes.

As future work, we are planning to work on the optimization of the efficiency of the system. Processing can be parallelized in order to take advantage of multiple processors executing the encrypted requests. We will also investigate how to reduce the number of recryptions needed. Indeed, since the noise value can be bounded, decryption should be necessary only when the ciphertext cannot support an additional arithmetic operation. We will are planning to develop a new scheme to encrypt the SQL circuits. In the current system, the server does know the operation that was performed (SELECT, UPDATE, etc.). If we can encrypt the SQL circuits, the system will preserve the confidentiality of the data and operations performed on these data. We believe that this new system can be the foundation of a highly secure cloud computing environment.

## References

[1] R. Rivest, A. Shamir, and L. Adleman, A Method for Obtaining Digital Signatures and Public-Key Cryptosystems, Communications of the ACM 21 (2): pp. 120–126, 1978.

[2] T. ElGamal, A Public-Key Cryptosystem and a Signature Scheme Based on Discrete Logarithms, IEEE Transactions on Information Theory, pp. 469–472, 1985.

[3] S. Goldwasser and S. Micali, Probabilistic Encryption. Journal of Computer and System Sciences, 28(2): pp. 270-299, April 1984.

[4] P. Paillier, Public-Key Cryptosystems Based on Composite Degree Residuosity Classes, Advances in Cryptology — EUROCRYPT '99 In Advances in Cryptology — EUROCRYPT '99, Vol. 1592 (1999), pp. 223-238, 1999.

[5] M. V. Dijk, C. Gentry, S. Halevi, and V. Vaikuntanathan, Fully Homomorphic Encryption over the Integers. EUROCRYPT 2010: pp. 24-43, June 2010.

[6] C. Gentry, Computing arbitrary functions of encrypted data, Commun. ACM, Vol. 53, No. 3., pp. 97-105, March 2010.

[7] C. Gentry, A fully homomorphic encryption scheme. PhD thesis, Stanford University, 2009.

[8] R. Chow, P. Golle, M. Jakobsson, R. Masuoka, and J. Molina, Controlling Data in the Cloud : Outsourcing Computation without Outsourcing Control. CCSW'09, pp. 85-90, Chicago, Illinois, USA, November 13, 2009.

[9] B. Chor, E. Kushilevitz, O. Goldreich, and M. Sudan, Private Information Retrieval, Journal of the ACM, 45(6): pp. 965-982, 1998.

[10] M. Raykova, B. Vo, and S. Bellovin, Secure Anonymous Database Search, CCSW'09, pp. 115-126, Chicago, Illinois, USA, November 13, 2009.

[11] N. Shang, G. Ghinita, Y. Zhou, and E. Bertino, Controlling Data Disclosure in Computational PIR Protocols. ASIACCS'10, pp. 310-313, Beijing, China, April 13–16, 2010.

[12] T. Nakamura, S. Inenaga, D. Ikeda, K. Baba, H. Yasuura, Anonymous Authentication Systems Based on Private Information Retrieval. Networked Digital Technologies. NDT '09, pp.53-58, 28-31 July 2009.

[13] S. Yinan and Z. Cao, Extended Attribute Based Encryption for Private Information Retrieval. Mobile Adhoc and Sensor Systems, 2009. MASS '09, pp. 702-707, 12-15 Oct. 2009.

[14] C. Gentry and Z. Ramzan, Single-Database Private Information Retrieval with Constant Communication Rate. ICALP 2005, LNCS 3580, pp. 803–815, 2005.

[15] C. A. Melchor and P. Gaborit, A Fast Private Information Retrieval Protocol. ISIT 2008, pp. 1848-1852, Toronto, Canada, July 6 - 11, 2008.